# Investigating Role of Personal Factors in Shaping Responses to Active Shooter Incident using Machine Learning

Ruying Liu, P.E.,[1] Burçin Becerik-Gerber, DDes.,[2] and
Gale M. Lucas, Ph.D.[3]

[1]Sonny Astani Department of Civil and Environmental Engineering, Viterbi School of Engineering, University of Southern California (USC), Los Angeles, CA; e-mail: ruyingli@usc.edu
[2] Dean's Professor, Sonny Astani Department of Civil and Environmental Engineering, Viterbi School of Engineering, USC, Los Angeles, CA.; e-mail: becerik@usc.edu
[3] USC Institute for Creative Technologies, USC, Los Angeles, CA; e-mail: lucas@ict.usc.edu

**ABSTRACT**

This study bridges the knowledge gap on how personal factors affect building occupants' responses in active shooter situations by applying interpretable machine learning methods to data from 107 participants. The personal factors studied are training methods, prior training experience, sense of direction, and gender. The response performance measurements consist of decisions (run, hide, multiple), vulnerability (corresponding to the time a participant is visible to a shooter), and pre-evacuation time. The results indicate that propensity to run significantly determines overall response strategies, overshadowing vulnerability, and pre-evacuation time. The training method is a critical factor where VR-based training leads to better responses than video-based training. A better sense of direction and previous training experience are correlated with a greater propensity to run and less vulnerability. Gender slightly influences decisions and vulnerability but significantly impacts pre-evacuation time, with females evacuate slower potentially due to higher risk perception. This study underscores the importance of personal factors in shaping responses to active shooter incidents.

**INTRODUCTION**

There has been a distressing escalation in the number of active shooter incidents (ASIs) in the United States in the past few years, presenting a significant threat to public safety and human life. A recent FBI report (FBI, 2023) highlights that in 2022 alone, there were 50 ASIs resulting in 100 fatalities and 213 injuries, representing a 66.7% increase from 2018. The escalation necessitates effective emergency response strategies. Predictive behavioral modeling for building occupants is essential for law enforcement and first responders to develop effective plans for such emergencies. Past studies have examined the decision-making processes and evacuation behaviors in indoor emergencies, particularly building fires. Existing evacuation simulation models typically focus on either interactions among individuals or interactions between individuals and the environment (Chen et al., 2021). For instance, predictive models have been used to estimate pre-evacuation behavior in fires, considering factors like responder group size and alarm cues (Zhao et al., 2020). Studies like Wang et al. (Wang et al., 2016) on near-field tsunami evacuation emphasize the significance of vertical evacuation structures in mortality reduction. D'Orazio et al. (D'Orazio et al., 2014) outlined the rules of motion for earthquake pedestrian evacuation, emphasizing the influence of environment configuration on route choice. Although there is comprehensive research on evacuation dynamics during natural disasters, the understanding of responses to human-caused emergencies, such as ASIs, is limited.

For ASIs, certain studies acknowledge the value of trained evacuation leaders for victim safety reflecting the influence of social factors (Arteaga et al., 2023), and the influence of environmental factors, such as security countermeasures (e.g., barriers, frosted windows, and staggering doors), on response time and decisions (Zhu et al., 2022). Yet, the correlation between human behaviors and personal factors in the context of ASIs has not been systematically explored. Indeed, in the context of natural disasters, personal factors such as gender, training experience, and spatial ability have been demonstrated to be fundamental in predicting human behaviors (Bateman & Edwards, 2002; Koshiba & Suzuki, 2018). For example, in response to hurricanes, females are more likely to evacuate due to gender roles, greater risk exposure, and higher risk perceptions (Bateman & Edwards, 2002). The significant correlation of personal factors with the emergency type highlights the need to examine these factors in specific emergencies, including ASIs.

In this paper, we answer the following research question: How do personal factors influence people's response behaviors in ASIs? We explore this research question by applying interpretable machine learning techniques, namely cluster analysis, decision tree, and linear regression, to predict evacuation decisions and performance metrics.

**METHODOLOGY**

This study utilizes data collected from the experiment assessing the efficacy of VR-based training for occupants' response and preparedness in ASIs (Liu et al., 2023a). Conducted in a controlled environment, the study isolated personal factors, eliminating the influence of environmental factors and social factors. The study was approved by the Institutional Review Board of the University of Southern California.

**Participants.** A total of 108 participants completed the experiment, with 54 males and 54 females. The analysis excluded one participant who consistently selected the same response across multiple questionnaires. Participants were undergraduate and graduate students with a mean age of 22.40 and a standard deviation of 3.32. Eligibility criteria included having a normal or corrected-to-normal vision, no heart-related illness, no mobility restrictions, and not experiencing sickness with VR equipment based on any prior exposure experience.

**Experimental Procedure.** Figure 1 illustrates the experimental procedure with a training session (label 2) and a testing session (label 4). The training session included a control group with video-based training and two immersive VR-based training groups (VR-low and VR-high) differing in interaction levels. The control group used a desktop computer to watch a video on the "run, hide, fight" strategy. The VR groups used a head-mounted display and Unity3D to create a virtual office setting. VR-low replicated the training content (a structured storyline comprising opening, 'run', 'hide', 'fight', and ending sections) in a virtual environment, allowing basic navigation with a controller. Built on the VR-low setting, VR-high group offered more interactivity and an adaptive storyline controlled by the experimenter using the Wizard of Oz technique, along with personalized feedback and recorded demonstrations of correct behaviors.

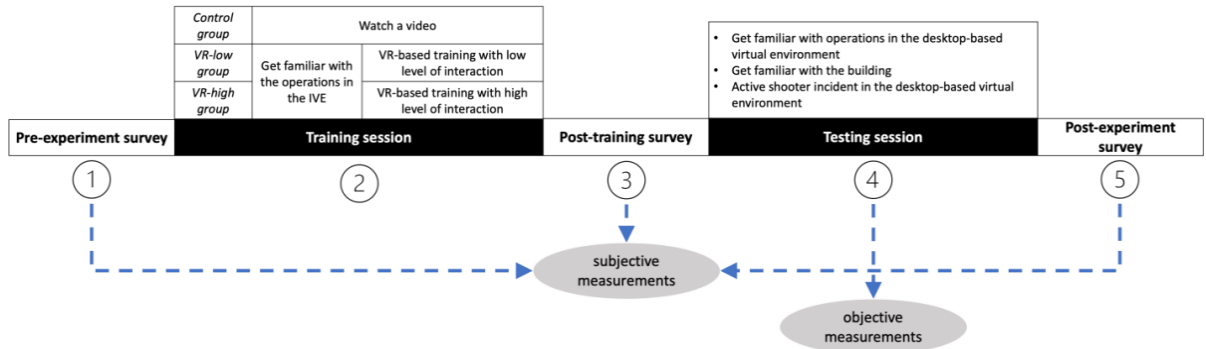

**Figure 1. Experimental procedure.**

To avoid bias in evaluating participants' response performance, the testing session (Label 4) used a desktop-based virtual environment, preventing any influence of VR familiarity on results. Participants interacted with the virtual environment using a mouse and keyboard, starting in a conference room with four avatars where a simulated gunshot occurred at the 10th second. The test concluded after 100 seconds. Subjective information was collected at three intervals: before, during, and after the experiment (Labels 1, 3, and 5 in Fig. 1). This study, detailed in (Liu et al., 2023a), compares VR-based and traditional training effectiveness for emergency responses, focusing on immersion, interactivity, and previous training experience. Utilizing the same dataset, this paper extends the scope to the exploration of personal factors affecting decisions and behaviors during ASIs.

**Feature Extraction.** To analyze participants' response performance in the testing session, we examined the metrics of vulnerability score, decision types (i.e., run, hide, multiple), and pre-evacuation time detailed in Table 1. The study identified training method as a key personal factor, established by randomly assigning participants to experimental groups. Alongside training method, factors such as gender, previous training experience, and spatial ability have been demonstrated to be fundamental in predicting human behaviors in the context of natural disasters (Bateman & Edwards, 2002; Koshiba & Suzuki, 2018; Thompson et al., 2017). In past experiments, these factors were often treated as confounding variables and balanced across groups to control their impact (Liu et al., 2023b). Consequently, the personal factors examined in this study include gender, training methods, previous training experience, and sense of direction as detailed in Table 2.

**Table 1. Description of collected response performance measurements.**

| Variables | Description |
|---|---|
| **Vulnerability** | Measured by the raycasting technique in Unity, correlates with the duration a participant is visible to a shooter, indicating the quality of a participant's response strategy. Increased by 5 points per second, ranging from 0 to 106. |
| **Decisions** | Categorized as run, hide, or multiple. There was no "fight" action. The "multiple" decision refers to the cases where participants altered their initial choices, resulting in multiple decisions throughout the testing scenarios. |
| **Pre-evacuation time** | Time taken (in seconds) to reach beyond the conference room doors from the moment of hearing an alarm. For those who hid in the conference room, it is the time they crouched at the hiding spot. The values range from 3 seconds to 45 seconds. |

**Table 2. Description of collected personal factors.**

| Variables | Description |
|---|---|
| **Gender** | Male/ Female |
| **Training method** | The type of training received before the simulation, which included the control group with video-based training, the VR-low group with low interactivity VR-based training, and the VR-high group with high interactivity VR-based training. |
| **Previous training experience** | Yes/No |
| **Sense of direction** | Environmental spatial ability assessed by the Santa Barbara sense of direction scale (SBSOD) (Hegarty et al., 2002), ranging from 1 (not at all) to 5 (very) |

**Data Processing.** Figure 2 shows our study's analytical framework, beginning with identifying dependent variables (Table 1) and using a two-step cluster analysis to categorize human response performance in ASIs. The analysis using the log-likelihood measure accommodates the mix of categorical and continuous variables in our dataset and is designed to discover inherent data groups. The two-step clustering process starts by forming pre-clusters to simplify the data matrix. Then, these pre-clusters are combined into final clusters using a hierarchical algorithm. The optimal number of clusters is automatically determined by Bayesian Information Criterion (BIC) or Akaike Information Criterion (AIC). This method resulted in two main clusters: 'Run' and 'Not Run,' where 'Not Run' includes hiding and multiple decisions.

The study employed decision tree analysis for its interpretability to examine decision-making based on clusters as the outcome variable. This method integrated independent variables of training methods, previous training experience, gender, and sense of direction. In preprocessing, categorical variables were numerically encoded. The data was split into 80% training and 20% testing subsets. The decision tree used the 'entropy' criterion for node splitting and was limited to three levels to prevent overfitting and maintain interpretability.

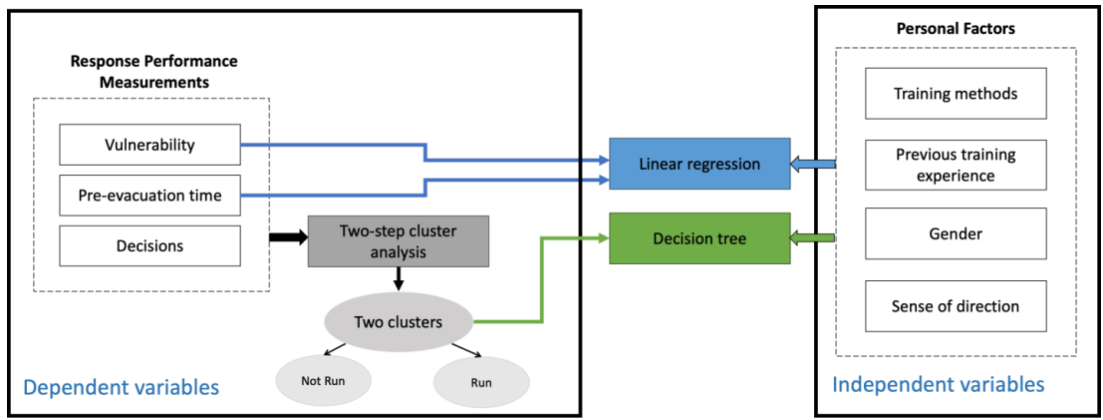

**Figure 2. System implementation plot for the analysis.**

Parallelly, linear regression was applied to the continuous dependent variables of vulnerability and pre-evacuation time to quantify the relationships with the same independent variables. One-hot encoding was applied to the group variable, creating binary columns for each experimental group. Continuous variables were normalized with MinMaxScaler, scaling them between 0 and 1, ensuring equal contribution from variables of different scales.

**Metrics for prediction assessment.** The cluster analysis quality was evaluated using the average silhouette score varying from -1 to 1, indicating cohesiveness and separation (Rousseeuw, 1987). Optimal clustering is characterized by minimal within-cluster distances and substantial between-cluster distances, aiming for a silhouette score near 1. A silhouette score over 0.5 denote high-quality clusters, while negative scores indicate poor clusters. The decision tree's performance was evaluated using accuracy and F1 score, with a baseline accuracy of 50% (i.e., 'Run' and 'Not Run') and a baseline weighted average F1 score of approximately 62.3% for random classification considering class imbalance. A decision tree model with both metrics exceeding the baseline suggests a performance better than random guessing. For linear regression, mean square error (MSE) and the coefficient of determination ($R^2$) were used to assess the results of the regression. Because achieving a high $R^2$ is challenging, a value larger than 0.1 can be acceptable for human studies (Ozili, 2023). Furthermore, given that understanding the model logic is as important as its predictive accuracy for our study, the feature importance of the cluster analysis and the decision tree were also interpreted, as well as the linear regression coefficients.

**RESULTS**

**Identification of Response Behavior Clusters (Run vs. Not run).** As shown in Figure 2, the two-step cluster analysis identified a 2-cluster solution with an average silhouette of 0.60, indicating a good cluster quality on participants' response performance measurements.

As summarized in Table 3, Cluster 1 (N = 80, 74.8%) corresponded to participants who took running decisions only, having lower vulnerability and shorter pre-evacuation time. Cluster 2 (N = 27, 25.2%) represented participants who did not make running decisions (i.e., hiding and multiple decisions instead), having higher vulnerability and longer pre-evacuation time. Among the three response performance parameters, the decision is the most important factor in estimating the model compared to vulnerability and pre-evacuation time.

Table 3. Cluster model summary.

| Input variables | Cluster 1 | Cluster 2 | Relative importance |
|---|---|---|---|
| **Decisions** | Run | Not run | 1 |
| **Vulnerability** | M = 10.04 | M = 26.89 | 0.14 |
| **Pre-evacuation time** | M = 10.13 | M = 14.20 | 0.12 |

**Predictors of Evacuation Decisions.** The decision tree model was visualized (see Figure 3) to illustrate the decision-making process for predicting running behavior in response to an active shooter incident. Table 4 shows the relative contribution of each feature to the model's decision-making process. The model achieved an accuracy of 86.36%, indicating a high rate of correct predictions for the binary classification of running behavior. Additionally, its weighted average F1 score of 85.57% suggests a strong balance of precision and recall across classes.

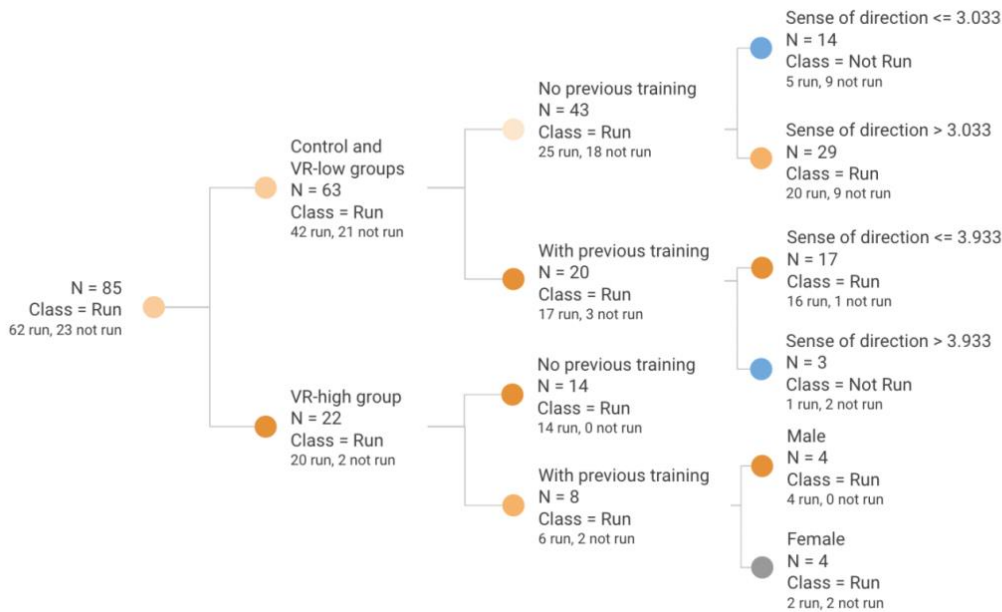

**Figure 3. The decision tree about the decisions of building occupants.**

The tree begins with the training method, suggesting its predominant influence on the initial decision to run, particularly differentiating immersive VR-based training with high interactivity from others. After this initial split, the tree branches based on previous training experience for both data subsets. It then further refines the classification using sense of direction for participants in less effective training methods (control and VR-low groups). For the VR-high group, the small subgroup size (N = 8) limits meaningful conclusions about gender influence. Also, sense of direction was not a significant factor in the VR-high group, suggesting its reduced impact compared to other training groups. The tree's final leaves display the classification outcome — either 'Run' or 'Not Run'.

The decision tree configuration indicates the dominance of training methods (i.e., VR-high group vs. other groups) and previous training experience at the top levels of the tree. The influence of the sense of direction in subsequent splits highlights its relevance to participants' decisions. Specifically, a higher sense of direction is associated with an increased propensity to run among first-time trainees within the control or VR-low groups. Conversely, a low sense of direction correlates with a tendency not to run. However, the inclination to run is consistent in the VR-high group, regardless of their previous training experiences.

**Table 4. Feature importance of factors influencing run/no-run behavior.**

| Personal factor | Feature importance |
| --- | --- |
| **Sense of direction** | 34.78% |
| **Previous training experience** | 32.85% |
| **Experimental group** | 20.09% |
| **Gender** | 12.28% |

**Factors Affecting Vulnerability.** The linear regression model investigating the influence of various personal factors on vulnerability exhibited an MSE of 349.50 with an $R^2$ of 0.20. Table 5 summarizes the regression coefficients for each personal factor.

**Table 5. Regression coefficients of personal factors influencing vulnerability.**

| Personal factor | Regression coefficient |
| --- | --- |
| **Sense of direction** | -8.75 |
| **Control group (video-based training)** | 8.43 |
| **VR-high group** | -1.32 |
| **Previous training experience** | -1.08 |
| **Gender (female)** | -0.26 |

*Note: The VR-low group is the reference category and is therefore not displayed in this table.*

The -8.75 coefficient for sense of direction indicates a strong negative relationship with vulnerability, suggesting individuals with better spatial awareness are less vulnerable. As stated in the data processing section where one-hot encoding was used to covert the experimental group variable into binary columns, the VR-low group was chosen as the reference category and therefore not displayed in Table 5. The coefficients of the control and VR-high groups represent the difference relative to the VR-low group. The positive 8.43 coefficient for the control group (i.e., video-based training) is associated with increased vulnerability compared to both VR groups, suggesting the effectiveness of VR-based training. Further, the negative coefficient -1.32 for the VR-high group, while smaller in magnitude, suggests that participation in the VR-high group decreases vulnerability compared to the VR-low group, underscoring the benefits of high interactivity VR-based training. The -1.08 coefficient for previous training experience indicates a slight vulnerability reduction, reinforcing the value of previous training, although its impact is less pronounced than other factors. Finally, the coefficient of -0.26 for gender implies a marginal negative relationship with vulnerability, with being a female slightly reducing vulnerability.

**Factors Affecting Pre-evacuation Time.** For investigating the influence of factors on pre-evacuation time, linear regression model exhibited an MSE of 13.77 with an $R^2$ of 0.30. Table 6 summarizes the regression coefficients for each personal factor.

**Table 6. Regression coefficients of personal factors influencing pre-evacuation time.**

| Personal factor | Regression coefficient |
|---|---|
| **VR-high group** | -2.85 |
| **Gender (female)** | 1.94 |
| **Control group (video-based training)** | 0.82 |
| **Previous training experience** | 0.05 |
| **Sense of direction** | -0.04 |

*Note: The VR-low group is the reference category and is therefore not displayed in this table.*

The -2.85 coefficient for the VR-high group suggests participants in this high-interactivity VR-based training evacuate quicker than others, indicating its effectiveness in active shooter response. Females, with a 1.94 positive coefficient, tend to have a longer pre-evacuation time than males. The positive coefficient of 0.82 for the control group indicates a slightly longer pre-evacuation time when compared to the baseline (i.e., VR-low group), indicating the influence of the lack of immersive training experience. Previous training experience, with a minimal 0.05 coefficient, has little effect on pre-evacuation time, highlighting its lesser importance relative to other factors. Finally, the small -0.04 coefficient for sense of direction indicates a marginally quicker evacuation decision for those with better spatial awareness.

**DISCUSSION**

Our findings indicate that the decision to run is the most critical predictor compared to other performance measures, such as vulnerability and pre-evacuation time. Personal factors (i.e., training methods, previous training experience, gender, and sense of direction) influence individual's response in active shooter situations. This aligns with literature, where these factors can significantly shape emergency response behaviors (Bateman & Edwards, 2002; Koshiba & Suzuki, 2018; Liu et al., 2023a; Thompson et al., 2017). Specifically, sense of direction is the most influential factor in predicting whether participants would run, with a 35% importance rate indicating that those with better sense of direction are more likely to take decisive actions. Sense of direction also exhibits a strong negative correlation with vulnerability, suggesting that effective navigators are less at risk. While a good sense of direction marginally speeds up evacuation, it has the smallest effect among the factors studied.

Previous training experience ranks just behind sense of direction in its influence on the decision to run, suggesting participants with such experience are more decisive, potentially due to increased preparedness. While its impact on vulnerability and pre-evacuation time is smaller, there is a slight decrease in vulnerability and a minor increase in pre-evacuation time associated with previous training. The training method also has a notable impact; VR-based training with high interactivity enhances running decisions, highlighting the value of immersive and interactive training. Participants in the control group show increased vulnerability and marginally longer pre-evacuation times, reflecting the limitations of video-based training in urgent decision-making scenarios. Gender holds the least importance among the factors regarding decision-making and vulnerability, which aligns with existing findings where no significant gender difference in fire evacuation behavior (Kinateder & Warren, 2016). However, females may take longer to begin evacuating, possibly due to heightened risk perception (Bateman & Edwards, 2002; Koshiba & Suzuki, 2018; Thompson et al., 2017).

While this study presents an attempt to employ interpretable machine learning to understand the influence of personal factors on responses to ASIs, it also has some limitations. It does not account for a wide range of personal factors, like age and ethnicity (Thompson et al., 2017), due to the homogeneity of the young adult student participants. Also, the training was received immediately before the testing simulation, which could overstate the influence of the training method without considering the potential long-term knowledge retention.

**CONCLUSION**

Our investigation advances an underexplored area of understanding the impacts of personal factors (e.g., training method, previous training experience, sense of direction, and gender) on occupants' response behaviors in ASIs (e.g., decisions, vulnerability, and pre-evacuation time). Machine learning analysis shows that the inclination to run significantly influences overall response strategies. Key factors such as sense of direction, prior training experience, and the type of training method received are critical in guiding these behaviors. While gender has a less pronounced effect on decision-making and vulnerability, it does contribute to differences in pre-evacuation time, where females generally evacuate more slowly possibly due to higher risk perception. These insights can improve safety protocols for first responders and training platforms for individuals to effectively respond to such emergencies.

**ACKNOWLEDGEMENTS**

This work was supported by National Science Foundation [Grant No. 1826443 and No. 2318559], Army Research Office [No. W911NF-20-2-0053], and USC Theodore and Wen-Hui Chen Endowed Fellowship. The views and conclusions contained in this paper are those of the authors and do not reflect the views of National Science Foundation or Army Research Office.

**REFERENCES**

Arteaga, C., Park, J., Morris, B. T., & Sharma, S. (2023). "Effect of trained evacuation leaders on victims' safety during an active shooter incident." *Safety Science*, *158*, 105967. https://doi.org/10.1016/j.ssci.2022.105967

Bateman, J. M., & Edwards, B. (2002). "Gender and Evacuation: A Closer Look at Why Women Are More Likely to Evacuate for Hurricanes." *Natural Hazards Review*, *3*(3), 107–117. https://doi.org/10.1061/(ASCE)1527-6988(2002)3:3(107)

Chen, J., Shi, T., & Li, N. (2021). "Pedestrian evacuation simulation in indoor emergency situations: Approaches, models and tools." *Safety Science*, *142*, 105378. https://doi.org/10.1016/j.ssci.2021.105378

D'Orazio, M., Spalazzi, L., Quagliarini, E., & Bernardini, G. (2014). "Agent-based model for earthquake pedestrians' evacuation in urban outdoor scenarios: Behavioural patterns definition and evacuation paths choice." *Safety Science*, *62*, 450–465. https://doi.org/10.1016/j.ssci.2013.09.014

FBI. (2023). *Active Shooter Incidents in the United States in 2022* [File]. https://www.fbi.gov/file-repository/active-shooter-incidents-in-the-us-2022-042623.pdf/view

Hegarty, M., Richardson, A. E., Montello, D. R., Lovelace, K., & Subbiah, I. (2002). "Development of a self-report measure of environmental spatial ability." *Intelligence*, *30*(5), 425–447. https://doi.org/10.1016/S0160-2896(02)00116-2

Kinateder, M., & Warren, W. H. (2016). "Social Influence on Evacuation Behavior in Real and Virtual Environments." *Frontiers in Robotics and AI*, *3*. https://www.frontiersin.org/articles/10.3389/frobt.2016.00043

Koshiba, Y., & Suzuki, Y. (2018). "Factors affecting post-evacuation behaviors following an earthquake: A questionnaire-based survey." *International Journal of Disaster Risk Reduction*, *31*, 548–554. https://doi.org/10.1016/j.ijdrr.2018.06.015

Liu, R., Becerik-Gerber, B., & Lucas, G. M. (2023a). "Effectiveness of VR-based training on improving occupants' response and preparedness for active shooter incidents." *Safety Science*, *164*, 106175. https://doi.org/10.1016/j.ssci.2023.106175

Liu, R., Zhu, R., Becerik-Gerber, B., Lucas, G. M., & Southers, E. G. (2023b). "Be prepared: How training and emergency type affect evacuation behaviour." *Journal of Computer Assisted Learning*, *n/a*(n/a). https://doi.org/10.1111/jcal.12812

Ozili, P. K. (2023). "The acceptable R-square in empirical modelling for social science research." In *Social research methodology and publishing results: A guide to non-native english speakers* (pp. 134–143). IGI Global.

Rousseeuw, P. J. (1987). "Silhouettes: A graphical aid to the interpretation and validation of cluster analysis." *Journal of Computational and Applied Mathematics*, *20*, 53–65.

Thompson, R. R., Garfin, D. R., & Silver, R. C. (2017). "Evacuation from natural disasters: A systematic review of the literature." *Risk Analysis*, *37*(4), 812–839.

Wang, H., Mostafizi, A., Cramer, L. A., Cox, D., & Park, H. (2016). "An agent-based model of a multimodal near-field tsunami evacuation: Decision-making and life safety." *Transportation Research Part C: Emerging Technologies*, *64*, 86–100. https://doi.org/10.1016/j.trc.2015.11.010

Zhao, X., Lovreglio, R., & Nilsson, D. (2020). "Modelling and interpreting pre-evacuation decision-making using machine learning." *Automation in Construction*, *113*, 103140. https://doi.org/10.1016/j.autcon.2020.103140

Zhu, R., Lucas, G. M., Becerik-Gerber, B., Southers, E. G., & Landicho, E. (2022). "The impact of security countermeasures on human behavior during active shooter incidents." *Scientific Reports*, *12*(1), Article 1. https://doi.org/10.1038/s41598-022-04922-8